\newcommand{\system}{\textsc{SIRD}\xspace} \newcommand{\globalB}{B\xspace}    \newcommand{\coreT}{NThr\xspace} \newcommand{\senderT}{SThr\xspace} \newcommand{\solT}{UnschT\xspace}  \newcommand{\overcommitname}{informed overcommitment\xspace} \newcommand{\Overcommitname}{Informed overcommitment\xspace} \newcommand{\wthree}{WKa\xspace} \newcommand{\wfour}{WKb\xspace} \newcommand{\wfive}{WKc\xspace} \newcommand{\data}{\texttt{DATA}\xspace} \newcommand{\grant}{\texttt{CREDIT}\xspace}

\newcommand{\sysRttSmIntra}{18\xspace}  \newcommand{\sysBdpSm}{216\xspace} \newcommand{\sysBdpSmFrames}{24\xspace}

\newcommand{\sysRttIntra}{\sysRttSmIntra}  \newcommand{\sysBdp}{\sysBdpSm} \newcommand{\sysBdpFrames}{\sysBdpSmFrames}

\newcommand{\killerbufvshoma}{12\xspace}    \newcommand{\killerbufvshomainsights}{13\xspace}

\newcommand{\killerutilvspim}{9}  \newcommand{\killerbufvspim}{43}   \newcommand{\killerlatvspim}{46}

\newcommand{\killerutilvsxpass}{26}   \newcommand{\killerlatvsxpass}{10\xspace}  

\newcommand{\vspacefigure}{}  

\documentclass[letterpaper,twocolumn,10pt]{article} \usepackage{usenix2019_v3} 

\usepackage{tikz} \usepackage{amsmath} \usepackage{subfloat} \usepackage{algorithm} \usepackage{algpseudocode} \algrenewcommand\textproc{} \usepackage[inline]{enumitem} \usepackage{varwidth} \usepackage{bm} \usepackage{balance} \usepackage{xcolor, soul} \usepackage{textcomp} \usepackage{graphicx} \sethlcolor{yellow}

\newif\ifcomments 

\usepackage{xspace}                           \usepackage[font=small]{caption} \usepackage{subcaption} \usepackage[normalem]{ulem}                  \usepackage{upgreek} \usepackage{comment}                         \usepackage{rotating} \usepackage{graphicx} \usepackage{multirow} \usepackage{multicol} \usepackage{gensymb}   \usepackage[htt]{hyphenat}    \usepackage{booktabs}

\usepackage[skip=0pt]{caption}

\makeatletter \@ifclassloaded{sigplanconf}{  }{     } \makeatletter

\newcommand{\printlayout}{{\bf The textwidth \the\textwidth; the columnwidth is \the\columnwidth}}

\newcommand{\eg}{\textit{e.g.,}\xspace} \newcommand{\ie}{\textit{i.e.,}\xspace}  \newcommand{\etal}{\textit{et al.}\xspace} \newcommand{\twiddle}{$\sim$} \newcommand{\microsecond}{$\upmu{}$s\xspace}  \newcommand{\nineninepct}{$99^{th}$ percentile\xspace}

 \newcommand{\myparagraph}[1]{\vspace{0.5em}\noindent {\bf #1:}}

\usepackage[english]{babel} \addto\extrasenglish{ \def\chapterautorefname{\S\@gobble} \def\sectionautorefname{\S\@gobble} \def\subsectionautorefname{\S\@gobble} \def\subsubsectionautorefname{\S\@gobble}   }

  

\begin{document} \pagestyle{empty}   

\date{}

\title{SIRD: A Sender-Informed, Receiver-Driven Datacenter Transport Protocol}

\author{ {\rm Konstantinos Prasopoulos}\\ EPFL
\and {\rm Ryan Kosta}\\ UCSD \thanks{Work done while at EPFL.}\\
\and {\rm Edouard Bugnion}\\ EPFL \and {\rm Marios Kogias}\\ Imperial College London }  

\maketitle 
\begin{abstract} 

Datacenter congestion control protocols are challenged to navigate the throughput-buffering trade-off while relative packet buffer capacity is trending lower year-over-year.  In this context, receiver-driven protocols --- which schedule packet transmissions instead of reacting to congestion --- excel when the bottleneck lies at the ToR-to-receiver link. However, when multiple receivers must use a shared link (\eg ToR to Spine), their independent schedules can conflict. 

We present \system, a receiver-driven congestion control protocol designed around the simple insight that single-owner links should be scheduled, while shared links should be managed with reactive control algorithms.  The approach allows receivers to both precisely schedule their downlinks \emph{and} to coordinate over shared bottlenecks. Critically, \system also treats sender uplinks as shared links, enabling the flow of congestion feedback from senders to receivers, which then adapt their scheduling to each sender's real-time capacity.  This results in tight scheduling, enabling high bandwidth utilization with little contention, and thus minimal latency-inducing buffering in the fabric. 

We implement \system on top of the Caladan stack and show that \system's asymmetric design can deliver 100Gbps in software while keeping network queuing minimal.  We further compare \system to state-of-the-art receiver-driven protocols (Homa, dcPIM, and ExpressPass) and production-grade reactive protocols (Swift and DCTCP) and show that \system is uniquely able to simultaneously maximize link utilization, minimize queuing, and obtain near-optimal latency. 

\end{abstract}



\section{Introduction} \label{sec:intro}

Datacenter workloads like ML training~\cite{cassini,nwaggrforml} and disaggregated resource management~\cite{networkreqsfordisag} increasingly demand high-throughput and low-latency networking.  While datacenter hardware continues to offer higher link speeds, switch packet buffer capacity is failing to keep up, and SRAM density trends show that this is unlikely to change~\cite{chang2022critical,deathofsram}.  The combination imposes a challenge for congestion control, which generally relies on buffering to deliver high throughput. The challenge is compounded by other important concerns such as hardware heterogeneity and cost of operations~\cite{vl2, ultraethernet}.

The most established approach to managing congestion is by reactively slowing down sender transmission rates after detecting a problem. This is the approach of sender-driven (SD) protocols~\cite{hpcc,dctcp,swift,powertcp,dcqcn,timely,hull,on-ramp,bolt,poseidon,fasttune, ppt} like DCTCP~\cite{dctcp} and Swift~\cite{swift} in which senders make decisions based on network feedback.   The reactive nature of SD protocols often requires several roundtrips to address congestion; a limitation for workloads dominated by small flows~\cite{homa,dcPIM,microscopic}.  Further, the flow abstraction makes it difficult to reduce latency through message-centric scheduling~\cite{pfabric,pias}. 

To address these limitations, a recent line of work~\cite{aeolus,homa,expresspass,ndp,dcPIM,fastpass,phost} proactively schedules packet transmissions instead of reacting to congestion buildup. Scheduling is typically implemented by having receivers send \emph{credit} packets to senders, which the latter consume to send back data. This enables tight control over ToR-to-Host ports which are usually the most congested~\cite{jupiter-rising, microbursts, microscopic}. The receiver-driven (RD) approach enables high throughput with limited buffering in dcPIM~\cite{dcPIM} and near-optimal latency in Homa~\cite{homa}. 

The fundamental tension in the design of receiver-driven schemes is how to schedule packets over shared links. Each receiver has exclusive control of its downlink but must share the bandwidth of the network's core and that of sender uplinks with other receivers. In a distributed protocol, this creates scheduling conflicts if receivers do not coordinate.  Proposed solutions for this problem include explicit pre-matching of senders and receivers (dcPIM~\cite{dcPIM}), in-network credit throttling (ExpressPass~\cite{expresspass}), and the overcommitment of receiver downlinks (Homa~\cite{homa}).   Despite impressive results, each of these approaches makes a sacrifice in either message latency, protocol complexity, or packet buffer utilization.

We propose \system, a receiver-driven design based on the following simple insight: exclusive links (receiver downlinks) should be managed proactively, and shared links (sender uplinks and switch-to-switch links) should be managed reactively. In practice, this means that receivers gather congestion feedback from shared links which gives them the means to coordinate, while still explicitly controlling their downlinks.  By deducing the bandwidth availability of shared links, and especially that of senders, receivers can be more precise in their credit allocation. This allows them to drive high link utilization without inducing contention and the corresponding queuing in the network. Low queuing, in turn, enables fast message delivery as the network does not introduce head-of-line blocking.

\system implements receiver-driven scheduling that reacts to shared-link congestion feedback using 1) end-host signals to detect congestion on sender uplinks, and 2) ECN~\cite{rfc3168} to detect congestion in the network's core. Congestion information flows via data packets to receivers and is fed into two independent control loops, with the most congested determining how much credit can be allocated to any given sender.  \system's design is end-to-end and does not require the use of in-network priority queues or unconventional switch configuration. 

We implement \system in 4300 SLOC within Caladan~\cite{caladan}, a state-of-the-art dataplane operating system. We show that \system can efficiently allocate credit, minimize queuing, and implement round-robin and SRPT policies at 100Gbps. 

We use network simulation to systematically compare \system to three state-of-the-art proactive protocols (Homa~\cite{homa}, dcPIM~\cite{dcPIM}, and ExpressPass~\cite{expresspass}) and two production-grade reactive protocols (DCTCP~\cite{dctcp} and Swift~\cite{swift}).  Our evaluation shows that \system can simultaneously maximize link utilization, minimize queuing, and obtain near-optimal latency.  Specifically: (1) \system causes $\killerbufvshoma\times$ less peak top-of-rack (ToR) buffering than Homa yet achieves competitive latency and utilization.  (2) Compared to dcPIM, \system has no message exchange rounds before sending and outperforms it in link utilization, peak ToR buffering, and tail latency by \killerutilvspim\%, \killerbufvspim\%, and \killerlatvspim\%.  (3) \system achieves $\killerlatvsxpass\times$ lower tail latency and $\killerutilvsxpass$\% higher utilization than ExpressPass. (4) \system outperforms DCTCP and Swift across the board, especially in incast-heavy scenarios.


\section{Background} \label{sec:back}

The main goal of congestion control (CC) is to allocate network resources in a work-conserving manner while avoiding packet loss. There are also other, at times conflicting, CC goals, \eg fairness, flow completion time or deadlines, and multi-tenancy.

Existing datacenter CC schemes for Ethernet networks can be broadly split into two categories: reactive and proactive. In reactive (also sender-driven (SD)) schemes~\cite{dctcp,hull,pfabric,on-ramp,swift,hpcc,timely,dcqcn,ppt}, senders control congestion by participating in a distributed coordination process to determine the appropriate transmission rates. Senders obtain congestion information through \emph{congestion signals} like ECN~\cite{rfc3168,dctcp, ppt}, delay~\cite{timely,swift}, and in-network telemetry~\cite{hpcc,poseidon}.

In proactive schemes~\cite{homa,phost,pias,ndp, expresspass, dcPIM, fastpass, pl2, harmony} bandwidth is allocated explicitly, either globally, or in a distributed manner.  Global approaches~\cite{fastpass, pl2}, use a centralized arbiter that controls all transmissions and hence face scalability challenges. Distributed proactive designs operate either end-to-end~\cite{homa, phost, pias, dcPIM} or with switch involvement in credit management~\cite{harmony, expresspass}. In end-to-end protocols, which are receiver-driven (RD), each receiver explicitly schedules its downlink by transmitting special-purpose credit tokens to senders. The credit rate can be explicitly controlled by a pacer~\cite{expresspass, ndp} or be self-clocked~\cite{homa,phost,dcPIM}.

\subsection{Exclusive and shared links} \label{sec:back:localizing}

Packets flowing to a receiver host can experience congestion either at the ToR-host link, which is exclusive to the receiver, or at host-ToR uplinks and the network core, which are shared among multiple receiving hosts.

\myparagraph{Exclusive links} In the context of RD protocols, ToR-to-Host links (downlinks) can be seen as exclusively controlled by a single entity, the receiver. By controlling the rate of credit transmission, a receiver can explicitly control the rate of data arrival - assuming the bottleneck is the downlink. In fact, downlinks are the most common point of congestion in datacenter networks~\cite{jupiter-rising, microbursts, swift, microscopic}. Congestion at the ToR downlink is the result of incast traffic from multiple senders to one receiver. In turn, incast is the result of the fan-out/fan-in patterns of datacenter applications~\cite{tail-scale}.

RD schemes excel in managing incast traffic because each receiver explicitly controls the arrival rate of data~\cite{homa,phost,pias,ndp, expresspass,dcPIM}. This level of control also allows RD schemes to precisely dictate which message should be prioritized at downlinks and can even factor-in application requirements~\cite{chimera}. Recent work has delivered significant message latency improvements by scheduling messages based on their remaining size (SRPT policy)~\cite{aeolus, homa, phost}. In contrast, SD schemes~\cite{dctcp,hull,pfabric,on-ramp,swift,hpcc,timely} treat downlinks as any other link. Senders must first detect incast at a receiver's downlink and then independently adjust their rates/windows such that the level of congestion falls below an acceptable target.

\myparagraph{Shared links} Unlike exclusive downlinks, core and sender-ToR link bottlenecks pose a challenge for RD schemes.  Shared link bottlenecks can appear when multiple receivers concurrently pull data packets over the same link.   Before discussing existing approaches to deal with this key challenge, we delve deeper into the specifics of shared-link congestion.

When congestion occurs in the network \emph{core}, multiple flows from unrelated senders and receivers compete for bandwidth on core links. Note that although the core may comprise multiple tiers, it is fundamentally shared infrastructure. Congestion at the core of a fabric is less common than at downlinks~\cite{homa,jupiter-rising,microscopic} but can still occur due to core network oversubscription. Oversubscription can be permanent to reduce cost~\cite{jupiter-rising} or transient due to component failures. Congestion at the core can also occur because of static ECMP IP routing decisions that cause multiple flows to saturate one core switch and one core-Tor downlink, while other core switches have idle capacity~\cite{vl2,conga}.

\emph{Uplink} congestion occurs due to the fan-out of multiple flows to different receivers or due to bandwidth mismatch between the sender's uplink and the receiver's downlink. Because the resulting packet buffering is in hosts and not in the fabric, sender congestion is a less severe problem from the perspective of packet loss.  For RD schemes, uplink congestion is known as the \textit{unresponsive sender problem}~\cite{homa} and leads to degraded throughput as independent scheduling decisions of receivers may conflict, wasting downlink bandwidth.  For example, two receivers may both decide to schedule the same sender $A$ at the same time, even if one of the receivers could be receiving data from sender $B$. Whereas the term \emph{unresponsive} has been used in the context of SRPT scheduling, where messages are transmitted to completion, we will use \textit{congested sender} as a general description of uplink congestion. 

RD protocols have proposed various mechanisms for overcoming the tension between independent receiver scheduling and the fact that some links are shared. One of the earliest designs, pHost~\cite{phost}, employs a timeout mechanism at receivers to detect unresponsive senders and direct credit to other senders. NDP~\cite{ndp} employs very shallow buffer switches and eagerly drops packets, using packet trimming to recover quickly. NDP receivers handle uplink sharing by only crediting senders if they transmit data packets - whether dropped or not.  pHost and NDP do not explicitly deal with core congestion and, further, analysis by Montazeri \etal~\cite{homa} showed that neither of the two achieves high overall link utilization. Homa~\cite{homa} introduced \textit{controlled overcommitment} in which each receiver can send credit to up to $k$ senders at a time. Homa achieves high utilization as it is statistically likely that at least one of the k senders will respond. However, it trades queuing for throughput by meaningfully increasing the amount of expected inbound traffic to each receiver. To let short messages bypass network queues, Homa leverages switch priority queues, which are typically used for application-level QoS guarantees~\cite{bolt, swift, hyperscaleissues}. dcPIM~\cite{dcPIM} employs a semi-synchronous round-based matching algorithm where senders and receivers exchange messages to achieve a bipartite matching.  This coordinates the sharing of sender uplinks but congestion at the core is only implicitly addressed by the protocol's overall low link contention. The downside of this link-sharing approach is message latency. dcPIM delivers small messages quickly by excluding them from the matching process. However, messages larger than the bandwidth-delay-product (BDP) of the network must wait for several RTTs before starting transmission. ExpressPass~\cite{expresspass} manages all links, exclusive and shared, via a hop-by-hop approach which configures switches to drop excess credit packets, which in turn rate limits data packets in the opposite direction. To reduce credit drops, ExpressPass uses recent credit drop rates as feedback to adjust the future sending rate, and in this way also improves utilization and fairness across multiple bottlenecks. Out of the designs discussed so far, ExpressPass is the only one that explicitly manages core congestion and can operate with highly oversubscribed topologies.  ExpressPass's hop-by-hop design helps it achieve near-zero queuing~\cite{expresspass} but is more complex to deploy and maintain due to its switch configuration and path symmetry requirements.   

\subsection{The impact of ASIC trends on buffering} \label{sec:asictrends}

Congestion control protocols depend on buffering to offset coordination and control loop delays and, as a result, face a throughput-buffering trade-off. Maximum bandwidth utilization can trivially be achieved with high levels of in-network buffering, but at the cost of queuing-induced latency and expensive dropped packet retransmissions. Conversely, low buffering can lead to throughput loss for protocols that are slow in capturing newly available bandwidth.

High-speed packet buffering is handled by small SRAM buffers in switch ASICs, the size of which is not increasing as fast as bisection bandwidth.    For example, nVidia's top-end Spectrum 4 ASIC has a $160$MB buffer, which corresponds to $3.13$MB per Tbps of bisection bandwidth~\cite{url:spectrum4}. The previous $12.8$Tbps and $6.4$Tbps top-end Spectrum ASICs were equipped with $5$MB and $6.6$MB per Tbps respectively~\cite{url:spectrum3,url:spectrum2} (full list in \autoref{app:appendix}). Unfortunately, future scaling of SRAM densities appears unlikely given CMOS process limitations~\cite{chang2022critical,deathofsram}. In parallel, datacenter round-trip times (RTT) are not falling as they are dominated by host software processing, PCIe latency, and ASIC serialization latencies.  Consequently, as link speeds increase, CC protocols must handle higher BDPs with less switch buffer space at their disposal.

To better absorb instantaneous bursts of traffic, switch packet buffers are generally shared among egress ports. Some ASICs advertise fully shareable buffers while others implement separate pools or statically apportion some of the space to each port~\cite{microscopic}.  On top of the physical implementation, various buffer-sharing algorithms dynamically limit each port's maximum allocation to avoid unfairness~\cite{dctcp}.  However, if a large part of the overall buffer is occupied, the per-port cap becomes more equitable, limiting burst absorbability~\cite{microscopic}.

\begin{figure}[t] \begin{minipage}{\linewidth} \begin{subfigure}{0.49\linewidth} \includegraphics[width=\linewidth]{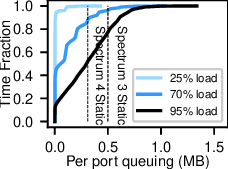}   \end{subfigure} \hfill \begin{subfigure}{0.49\linewidth} \includegraphics[width=\linewidth]{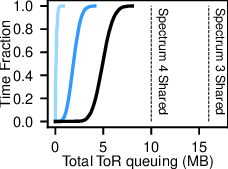}    \end{subfigure} \end{minipage}

\caption{ Homa queuing CDFs under various network loads for workload Websearch~\cite{pfabric}. The dotted lines represent the switch buffer size adjusted to the actual radix of our simulated ToR; see \autoref{sec:eval-methodology}.  } \label{fig:buffering-cdf}  \end{figure}

\autoref{fig:buffering-cdf} provides some context by comparing the buffer use of Homa, a state-of-the art RD protocol, with recent switch buffer capacities under the Websearch workload~\cite{pfabric}.          The dotted lines represent the per-port (left) and completely shared (right) buffer capacities of the hypothetical switches, adjusted on a per-unit basis to the port radix of our simulation methodology (\autoref{sec:eval-methodology}).  In practice, the buffer is neither partitioned nor fully shared; however, between the two extremes, hardware trends are constraining the throughput-buffering trade-off.

\section{\system Design Pillars} \label{sec:insights}

\system is an end-to-end receiver-driven scheme that manages exclusive links proactively and shared links reactively. In practice, this means that receivers perform precise credit-based scheduling when the bottleneck is their downlinks and use congestion feedback to coordinate over shared links.  \system does not face the limitations of prior work as it:

\begin{itemize}[topsep=1pt] \setlength{\itemsep}{1pt} \setlength{\parskip}{0pt} \item is end-to-end, with all decision-making happening at end hosts, and does not rely on advanced switch features or configuration. \item achieves high throughput by using sender and core link congestion feedback to allow receivers to direct credit to links with spare capacity. \item causes minimal buffering in the network by drastically reducing the need for downlink overcommitment.   \item does not need switch priority queues to deliver messages with low latency, thanks to minimal buffering. \item can start message transmission immediately without a prior matching stage. \item explicitly tackles core congestion through its shared link management approach. \end{itemize}

\myparagraph{Efficient credit allocation}  \system's design allows receivers to send the appropriate amount of credit depending on the real-time availability of the bottleneck link. For downlinks the task is trivial as the link's capacity is managed by one receiver. However, when the bottleneck is shared (sender uplink or core link), \system receivers detect competition for bandwidth and adjust the amount of issued credit dynamically. For example, if a sender is the bottleneck for two receivers, each will allocate the appropriate amount of credit for their share of the sender's bandwidth.  This approach makes the distribution of credit efficient because it is given to hosts that can promptly use it rather than being accumulated by congested senders.  As a result, \system can achieve high link utilization using a small amount of credit, \ie with little overcommitment.  

\system combines efficient credit allocation and overcommitment in \emph{\overcommitname}.   Each receiver is allotted a limited amount of available credit $\globalB$. The minimum valid value of $\globalB$ is $1\times BDP$ as this is the amount of credit required to pull $1\times BDP$ of inbound traffic and fully utilize the downlink. Higher values of $\globalB$ lead to downlink overcommitment, which increases tolerance to congested shared links but also increases expected queuing. 

\begin{figure} \centering\includegraphics[width=\linewidth]{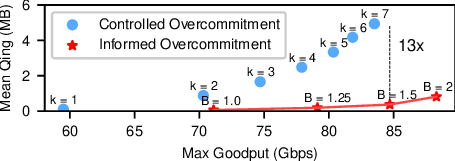} \caption{Mean buffering at ToRs versus maximum achieved goodput when sweeping the overcommitment parameter for \system (\overcommitname) and Homa (controlled overcommitment).  Results obtained in simulation by running the Websearch workload at 95Gbps on 100Gbps links across 144 servers;  see \autoref{sec:eval-methodology}. }

\label{fig:frontier} \vspacefigure

\end{figure}


\autoref{fig:frontier} highlights the benefit of efficient credit allocation in \system by comparing the throughput-buffering trade-off across equivalent overcommitment levels for Homa ($k$), which introduced and uses RD overcommitment, and \system ($\globalB$) under maximum network load. \Overcommitname yields the same overall goodput despite overcommitting downlinks $14\times$ less, which leads to $\killerbufvshomainsights\times$ lower queuing. The goodput-queuing trade-off benefit is maximum at high network load, under which congestion and loss are most likely, and naturally fades as load lessens.

\system implements \overcommitname through a control loop at each receiver, which dynamically adjusts credit allocation across senders based on sender and network feedback. When a sender is concurrently credited by multiple receivers, the sender receives credit faster than it can consume it, causing it to accumulate.  This is undesirable as receivers have a limited amount of credit to distribute.   Equivalently, limiting accumulation makes credit available to other senders that can actually use it. \system's control loop rebalances credit by setting an accumulated credit threshold, $\senderT$, that is similar in spirit to DCTCP's marking threshold~\cite{dctcp}, Swift's target delay\cite{swift}, or HPCC's $\eta$~\cite{hpcc}. At each sender, while the amount of accumulated credit from all receivers exceeds $\senderT$, a bit is set in all outgoing data packets. Based on arriving bit values, each receiver adjusts the maximum amount of credit that can be allocated to the congested sender in an additive-increase/multiplicative-decrease (AIMD) manner. In this paper, we use DCTCP's~\cite{dctcp} AIMD algorithm. 

\system uses the same mechanisms to handle core congestion with ECN as the signal.  Receivers read the CE bit of data packets and adjust per-sender credit limits using AIMD. 

Each receiver runs two AIMD algorithms in parallel, one for senders and one for the core, each with its own input signal. The algorithm instance of the most congested area determines credit allocation.   This separation of signals allows \system to use different signal/algorithm combinations to handle bottlenecks at hosts and switches.  Beyond ECN, \system can use signals such as end-to-end delay on infrastructures with timestamping support~\cite{swift} or In-Band Telemetry~\cite{hpcc, poseidon}.   In this paper, we use the same proven and simple combination for both and leave the exploration of more complex algorithms and signals to future work.

\myparagraph{Starting at line rate} Prior work~\cite{homa, phost,aeolus,ndp} has demonstrated the latency benefits of starting transmission at line rate, without a gradual ramp-up or preceding control handshakes that take at least one RTT. \system's credit allocation mechanism operates \emph{while transmission progress is made}, and thus, senders immediately start at line rate, sending the first BDP bytes \textit{unscheduled} (without waiting for credit). 

We further optimize the design based on the following simple observation: small messages benefit the most from unscheduled transmission since their latency is primarily determined by the RTT while throughput-dominated messages see minimal gain. For example, in the absence of queuing, delaying transmission by one RTT increases the end-to-end latency of a message sized at $10\times BDP$ by 9\% compared to 200\% for a single-packet message. Therefore, to reduce unnecessary bursty traffic and queuing, \system senders do not start transmitting messages larger than a configurable threshold (\solT) before explicitly receiving credits from the receiver.  For messages smaller than \solT, senders send the first $min(BDP, msg\_size)$ bytes without waiting for credit.   

\myparagraph{Switch priority queues not required} By efficiently allocating credit, \system achieves high link utilization with minimal fabric queuing, thereby largely eliminating its impact on latency. Thus, \system can be deployed without any in-network QoS support, though it can benefit in terms of tail latency from having a total of two priority levels. 


\section{\system Design}

\label{sec:design}

At a high level, \system is an RPC-oriented protocol, similar to other recently proposed datacenter transports~\cite{homa,phost,dcPIM}. \system may be used to implement one-way messages or remote procedure calls~\cite{homa,erpc,r2p2}. \system assumes that the length of each message is known or that data streams will be chunked into messages. \system further assumes that ECN is configured in all network switches, with the ECN threshold set according to DCTCP best practices~\cite{dctcp}.   \system is designed to be layered on top of UDP/IP for compatibility with all network deployments. The UDP source port is randomly selected for each packet for fine-grain load balancing, allowing an ECMP network to behave as effectively as random packet spraying~\cite{random-packet-spraying}. We make no assumption of lossless delivery, in-network priority queues, or smart NICs and switches. 

\system defines two main packet types: {\bf (1)} \data: a packet that contains part of the payload of a message.   Data packets can be scheduled, requiring credit to be sent, or unscheduled. Messages with size $> \solT$ are scheduled and make an initial credit request by sending a zero-length \data packet;     {\bf (2)} \grant: a control packet sent by the receiver to the sender to schedule the transmission of scheduled \data.   

Generally, the flow of packets consists of credit flowing from receivers to senders and data flowing in the opposite direction. Credit leaves the receiver in a \grant packet, is used by the sender, and returns with a scheduled \data packet.

\subsection{Credit Management} \label{sec:credit-management}

Each \system receiver maintains two types of credit buckets that limit the amount of credit it can distribute: a global credit bucket and per-sender credit buckets.

The \emph{global credit bucket $\globalB$} controls overcommitment by capping the total number of outstanding credits that a receiver can issue. Configuring $\globalB \geq BDP$ is necessary to enable link saturation. Further, $\globalB$ bounds the queuing length from scheduled packets to $\globalB-BDP$ bytes at the ToR's downlink. Our analysis (\autoref{sec:informed-overcommitment}) and evaluation (\autoref{sec:eval}) show that setting $\globalB$ as low as $1.5\times BDP$ is sufficient for high link utilization. 

Each receiver also maintains a credit bucket per sender machine it communicates with. The \emph{per-sender credit bucket} caps the number of outstanding credits the receiver can issue to a sender.  \Overcommitname is implemented by adjusting the size of the per-sender credit bucket according to the level of congestion at the core and the sender (max $1\times BDP$). Reducing the bucket size means that less of the receiver's total credit can be allocated to a congested source or path, and thus, more is available for other senders.

\subsection{Informed Overcommitment} \label{sec:informed-overcommitment}

\Overcommitname uses two input signals that communicate the extent of congestion in the network and at senders. The signals are carried in \data packets and are the ECN bit in the IP header set by the network and a bit in the \system header set by the sender. Each receiver runs two separate AIMD (additive-increase, multiplicative decrease) control loops and uses the most conservative of the two (similar to Swift~\cite{swift}) to adjust per-sender credit bucket sizes. Each control loop is configured with its own marking threshold ($\coreT$ and $\senderT$). $\coreT$ should be set according to DCTCP best practices~\cite{dctcp} to limit queuing in the network core.  Note that $\coreT$ is much higher than the maximum allowed $\globalB - BDP$ of persistent queueing at ToR switches. Thus, ToR switches never have to mark ECN.

$\senderT$ should be set to limit the amount of credit a sender can accumulate, thus allowing receivers to efficiently distribute their limited aggregate credit. Intuitively, $\senderT$ determines the level of accumulated credit the control loop is targeting when a sender is congested, \ie receiving credit faster than it can use it. Setting it too high means each sender can accumulate substantial amounts of credit, and consequently, $\globalB$ would need to be configured higher to increase aggregate credit availability. Conversely, setting $\senderT$ too low does not allow the control loop any slack when converging to a stable state and can cause throughput loss.

To understand the relationship between $\senderT$ and $\globalB$, we analytically examine a simple congested-sender scenario from the perspective of a receiver $R$ to find: how much total credit ($\globalB$) does $R$ need when receiving from $k$ congested senders, to still have enough credit available to saturate its downlink. Assuming each congested sender sends to $f$ receivers in total, its available uplink bandwidth for $R$ is $BW/f$. Assuming uniform link speeds, we are interested in the case where senders are the bottleneck and cannot saturate $R$, or:

\begin{equation} BW_{supply\_k} < BW_{demand\_R} => \frac{BW}{f} < \frac{BW}{k} => f > k \label{equation:bottleneck} \end{equation}

In this case, each of the $k$ senders accumulates up to $\senderT$ credit in stable state, or $\senderT / f$ from each receiver assuming an equal split (see~\autoref{sec:pacing-scheduling}).  Therefore, to be able to saturate its downlink, $R$'s $B$ must be large enough to allow $1xBDP$ of credit to be in flight despite accumulation at congested senders, or:       

\begin{equation} B \geq BDP + \sum_k{\frac{\senderT}{f} \;;\;  f \geq 2, f > k}   \label{equation:htr} \end{equation}

If $k_{max}$ is the number of congested senders that maximizes the right side of the inequality, then $f_{max} = k_{max}+1$ ($f$ in denominator) and the maximum value of the term is:

\begin{equation} \sum_{k}^{k_{max}}{\frac{\senderT}{k_{max}+1}} = \frac{\senderT}{k_{max}+1} \sum_{k}^{k_{max}}{1} = \senderT \frac{k_{max}}{k_{max}+1} < \senderT \label{equation:max} \end{equation} 

It follows that, in steady state, $R$ can account for any number of congested senders as long as $B \geq BDP + \senderT$.  Under dynamic traffic patterns (see \autoref{sec:eval-sensitivity}) higher values of B can help increase the supply of credit in transient phases. Further, policies other than fair sharing can loosen this property. For example, if senders strictly prioritize some receivers, then, in the worst case where $R$ is de-prioritized by all k senders, \autoref{equation:htr} loses $f$ from the denominator and B depends on the number of worst-case congested senders.


\begin{table}[b]  \caption{Core configuration parameters.} \centering \begin{tabular}{ |c | p{2.5in}| } \hline  
\multirow{2}{*}{\solT }  & Messages that exceed \solT in size ask for credit before transmitting.\\
\hline
\multirow{2}{*}{\globalB }  & Per-receiver global credit bucket size. Caps credited-but-not-received bytes.\\
\hline
\multirow{1}{*}{\coreT } & ECN threshold, configured as for DCTCP.\\
\hline
\multirow{1}{*}{\senderT } & Sender marking threshold ($sird.csn$). \\
\hline \end{tabular} \label{table:parameters} \end{table}



\begin{figure}[t]  \begin{algorithm}[H] \footnotesize \caption{Receiver Logic}\label{alg_receiver}

\textbf{Variables:} \begin{itemize}[topsep=0pt, itemsep=0pt]   \item $\bm{b}$: consumed credit from global receiver bucket of size \globalB,  \item $\bm{sb_i}$: consumed credit from the bucket of sender i,  \item $\bm{senderBkt_i, netBkt_i}$: Sender and network credit bucket size, \item $\bm{rem_i}$: Requested but not granted credit for sender i.   \end{itemize} 

\begin{algorithmic}[1]

\Procedure{onDataPacket}{$pkt,i$}\Comment{i:sender} \State $credit \gets \Call{getCredit}{pkt}$  \State $b \gets b - credit$ \State $sb_i \gets sb_i - credit $ \State $senderBkt_i \gets \Call{Sender_{AIMD}}{senderBkt_i,pkt.sird.csn}$ \State $netBkt_i  \gets \Call{Net_{AIMD}}{netBkt_i,pkt.ip.ecn}$ \EndProcedure

\vspace{1mm}

\Procedure{onSendCreditTick()}{} \Comment{Runs when $b + min(rem_i,MSS) \leq \globalB$} 

\State \begin{varwidth}[t]{\linewidth} $senderList \gets activeSenders.filter($\par \hskip\algorithmicindent $sb_i + min(rem_i,MSS) \leq min(senderBkt_i,netBkt_i))$ \end{varwidth} \State $s \gets \Call{policySelect}{senderList}$ \State $credit \gets min(rem_i,MSS)$ \State $\Call{sendCredit}{s,credit}$ \State $b \gets b + credit$; $sb_i \gets sb_i + credit$; $rem_i \gets rem_i - credit$ \EndProcedure \end{algorithmic} \end{algorithm}   \end{figure}

\subsection{Congestion Control Algorithm} \label{sec:core-design}

Algorithm~\ref{alg_receiver} describes the behavior of a \system receiver.  When credit is available in the global bucket and on the command of the credit pacer (\autoref{sec:pacing-scheduling}), the receiver tries to allocate credit to an active sender (ln.~8). It first selects one of the senders with available credit in the per-sender bucket (ln.9) based on policy (ln.10).  The receiver sends a \grant packet (ln.~12) and reduces the available credit in the global and per-sender buckets, and updates the remaining required credit for that message (ln.~13).   Whenever a \data packet arrives (ln.~1), if it is scheduled, the receiver replenishes credit in the global bucket (ln.~3) and the per-sender bucket (ln.~4). Then, it executes the two independent AIMD control loops to adapt to a congested sender (ln.~5) and a congested core network (ln.~6), respectively. The receiver sets the size of the per-sender bucket as the minimum of the two values (ln.~9). 

\begin{figure}  \begin{algorithm}[H] \footnotesize \caption{Sender Logic}\label{alg_sender} \textbf{Variables:} $\bm{c_r}$: available credit for outbound messages to receiver r \begin{algorithmic}[1] 

\vspace{1mm}

\Procedure{onCreditPacket}{$r, credit$} \State $c_r \gets c_r + credit$ \EndProcedure

\vspace{1mm}

\Procedure{sendData()}{} \State $rcvrList \gets activeReceivers.filter(c_r > 0)$; \State $(r,dataPkt) \gets \Call{policySelect}{rcvrList}$  \State $dataPkt.sird.csn \gets (\sum{_i}(c_i) \geq \senderT)$ \State $c_r \gets c_r - dataPkt.size$ \State $\Call{send}{r,dataPkt}$ \EndProcedure

\end{algorithmic} \end{algorithm}   \end{figure}


Algorithm~\ref{alg_sender} describes the implementation of the sender-side algorithm for scheduled \data packets.  A host can send data to a receiver only if it has credit from said receiver.        Congested senders mark the \emph{congested sender notification (sird.csn)} bit if the total amount of accumulated credit exceeds $\senderT$ (ln.~7). The sender can send unscheduled \data packets at any point in time.   It is worth noting that \system senders naturally handle scenarios where a meaningful portion of uplink bandwidth is consumed by unscheduled packets. Since senders limit credit accumulation by informing receivers to allocate less credit, the transmission rate of scheduled packets converges to the leftover bandwidth.  

\subsection{Other Design Concerns} \label{sec:pacing-scheduling}

\system can be configured to schedule for fairness \eg by crediting messages in a round-robin manner, or for latency minimization, \eg by crediting smaller messages first (SRPT), or to accommodate different tenant classes.   \system implements policies at the receiver (ln~.10), which is the primary enforcer, and at the sender (ln.~6). By minimizing queuing in the fabric, \system does not need to enforce policies there, simplifying the design.  Regardless of which policy is configured at senders, \system allocates part of the uplink bandwidth fairly across active receivers, as to ensure a regular flow of congestion information between sender-receiver pairs.

\system receivers pace credit transmission to match their downlink's capacity.  Pacing improves message latency by reducing downlink queuing from scheduled packets even below the tight $\globalB - BDP$ bound but is not needed for correctness.

If switch priority queues are available, \grant packets are sent over a higher priority lane to further reduce RTT jitter. The unscheduled prefixes of messages smaller than $\solT$ also use this lane to bypass transient network queueing.    \system does not require in-network priority queues to deliver high performance but sees tail latency benefits in some cases (\autoref{sec:eval-sensitivity}).  

Packet loss in \system will be very rare by design, but the protocol must still operate correctly in the presence of CRC errors or packet drops due to faults or restarts. \system employs Homa's~\cite{homa} retransmission design in which receivers infer loss when no new packets are received for an incomplete message after a period of a few milliseconds. Upon detecting the loss of a scheduled segment, receivers also reclaim the credit allocated for said segment. \system's modest credit overcommitment (\eg 50\%) ensures that receivers can continue fully utilizing their bandwidth even in the improbable event where a sizeable chunk of packets is lost. Finally, \system incurs no performance penalties from out-of-order packet deliveries as it does not rely on packet order for loss detection and only delivers completed messages to applications. 


\section{Caladan Implementation} \label{sec:impl}

\system, with its message-level API and simple credit management design, can be implemented in NIC hardware or in software. The former eliminates PCIe latencies from the RTT, reducing the $BDP$, and allows for nanosecond-scale pacing and delivery of packets on the wire.   We demonstrate that the latter is also possible, at 100Gbps, by implementing \system in 4300 SLOC as a UDP-based transport protocol in Caladan~\cite{caladan}, a state-of-the-art kernel-bypass dataplane operating system and CPU scheduler.     We chose Caladan because of its mature network stack and its lightweight green thread abstraction, suitable for building latency-sensitive applications. \system is open source and available on GitHub~\cite{sird-implementation}.

The \system stack follows an asymmetric design where dedicated spinning threads perform specialized tasks. On the receiver-side, each \system node has one dedicated thread that runs the control loops and implements downlink scheduling policies, like round-robin or SRPT, by transmitting paced \grant{s} according to Algorithm~\ref{alg_receiver}.  \system reduces network queuing by pacing credit at slightly less than the line rate, as proposed in Hull~\cite{hull}.   Dedicated receiver cores read from RX rings and adjust credits, ensuring the system can quickly replenish new credits for redistribution. Each dedicated receiver core has two RX rings, one for packets of scheduled messages and one for packets of unscheduled messages; this helps avoid drops of scheduled packets during high rates of unscheduled traffic. For efficiency, and to sustain 100Gbps line rates, memory allocation and data copying operations are removed from the credit replenishment path to a small set of dedicated threads which copy packets, reassemble messages from packets, which can be received out of order, and deliver messages to applications via inline callbacks. 

The sender-side implementation also has a central thread, which accepts \grant{} packets and schedules the transmission of all \data packets, whether these are unscheduled or scheduled.  This thread implements Algorithm~\ref{alg_sender} which marks the $sird.csn$ bit of \data packets by comparing the amount of accumulated credit to $\senderT$.   Like the central receiver thread, this thread paces itself to ensure that packets are not given to the NIC faster than it can consume them. The thread accepts new message requests via asynchronous shared-memory queues, where the application API returns error codes if resource limits are exceeded.  



\section{Evaluation} \label{sec:eval}

We evaluate \system using our Caladan implementation and through simulation to answer the following questions: {\bf(1)} Can a software implementation of \system deliver 100Gbps along with minimal queuing under incast (\autoref{sec:evalimpl:incast}) and efficient credit management under outcast (\autoref{sec:evalimpl:outcast})? {\bf(2)} How well does \system navigate the throughput-buffering-latency trade-off in larger scale workloads compared to existing work (\autoref{sec:Throughput-Buffering-Latency})?   {\bf(3)} Is \system's congestion response robust at high load pressure and can it handle core congestion (\autoref{sec:Congestion-Response})?   {\bf(4)} Can \system deliver messages with low latency in large scale workloads (\autoref{sec:message-latency})? {\bf(5)} How important is \system's sender-informed design in maximizing utilization with minimal buffering? How sensitive is \system to its parameters (\autoref{sec:eval-sensitivity})?

\subsection{System Evaluation} \label{sec:evalimpl}

We evaluate \system at a small scale using Cloudlab's~\cite{cloudlab} sm110p machines equipped with 100Gbps ConnectX-6 DX NICs and Xeon 4314 CPUs, all located in the same rack. Our implementation uses 9KB jumbo frames and the unloaded RTT of the \system{}-Caladan stack is approximately \sysRttIntra\microsecond. Latency measurements are end-to-end, measured by the client, and include four request/reply copies to/from the application layer. We sample other system metrics every 2ms. While the size of requests may vary from experiment to experiment, replies are of minimal size.

We configure \system parameters as follows: $BDP=\sysBdp KB$ (\sysBdpFrames jumbo frames), $\globalB=1.5\times BDP$, $\senderT=0.5\times BDP$, $\solT = 1.0\times BDP$. We do not use switch priority queues.

\subsubsection{Receiver Congestion} \label{sec:evalimpl:incast}

\system's main objectives when handling incast are eliminating queuing-induced latency for small messages and being able to reduce latency for larger messages through credit-based scheduling.  To evaluate this, we run an experiment where six senders saturate a receiver by sending 10MB requests at a rate of 17Gbps each. A seventh sender periodically transmits either 8B or 500KB requests and captures the latency distribution.

\begin{figure}     

\begin{subfigure}{0.49\linewidth} \includegraphics[width=\columnwidth]{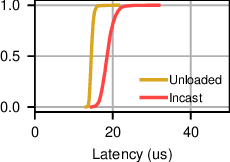} \end{subfigure} \hfill \begin{subfigure}{0.49\linewidth} \includegraphics[width=\columnwidth]{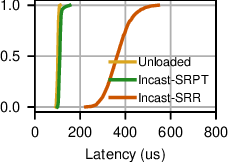} \end{subfigure}

\caption{ {\bf Incast:} CDF of message latency under incast compared to an unloaded baseline for 8B requests (left) and 500KB requests (right). The incast is formed by six senders transmitting 10MB messages in an open loop.  } \label{fig:syseval-incast1} \vspacefigure  \end{figure}


\autoref{fig:syseval-incast1} (left) plots the round-trip latency experienced by short requests, which are unscheduled. When the receiver is saturated, we observe only a few microseconds of additional latency compared to when the receiver is unloaded, which corresponds to a couple of packets of queuing, either at the ToR or at the host stack. For context, the median latency under load for the same experiment using kernel TCP Cubic was above 1ms.   \system achieves this result because it limits the number of scheduled inbound bytes to $B$ and thus the extent of downlink queuing to $B-BDP$. Further, since in this case senders respond to credit immediately, credit pacing effectively eliminates downlink queuing, while still achieving 96Gbps in this experiment.  

\autoref{fig:syseval-incast1} (right) shows the latency of \emph{scheduled} 500KB requests under two receiver scheduling policies.   Under SRPT, the receiver allocates credit to the message with the fewest remaining bytes and thus quickly prioritizes 500KB over 10MB requests, leading to near-unloaded latency despite the saturated downlink. \system can also implement fairer policies like per-sender round robin ("SRR") as shown. Finally, as it causes minimal network queuing, \system can implement such policies without in-network QoS support.

\subsubsection{Sender Information} \label{sec:evalimpl:outcast}

\begin{figure}

\begin{subfigure}{\linewidth} \centering \includegraphics[width=0.6\linewidth]{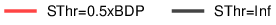} \end{subfigure} \vspace*{-1em}

\begin{subfigure}{0.49\linewidth} \includegraphics[width=\linewidth]{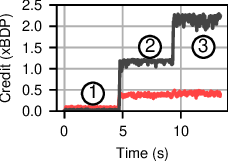} \end{subfigure} \hfill \begin{subfigure}{0.49\linewidth} \includegraphics[width=\linewidth]{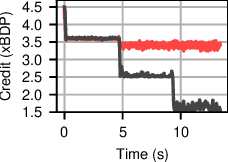} \end{subfigure}

\caption{Left: credit accumulated at congested sender. Right: sum of credit available at the three receivers - initial total: $4\times 1.5 = 4.5\times BDP$. 100ms moving average. The circled numbers indicate the number of receivers at that stage of the experiment.} \label{fig:syseval-outcast1}   \end{figure}

When senders are the bottleneck (outcast), \system tackles the congested sender problem by providing congestion feedback to receivers and allowing them to adjust their credit allocation. To evaluate the mechanism's efficacy at 100Gbps, we run an outcast experiment where a single sender sends 10MB messages at full rate to three receivers in a time-staggered manner. \autoref{fig:syseval-outcast1} (left) shows that, without \overcommitname ("$\senderT=Inf$"), the level of credit accumulated at the congested sender increases with every new receiver. This is because each receiver independently gives the sender $1\times BDP$ worth of credit to achieve full rate. Through congestion feedback ("$\senderT=0.5\times BDP$"), receivers coordinate and scale down their credit allocation until credit accumulation at the sender is less than $\senderT = 0.5\times BDP$, on average. Credit, instead, stays at receivers, and can be used to schedule other senders. \autoref{fig:syseval-outcast1} (right) shows the sum of credit available at the three receivers, each receiving a max-min fair share of the sender's bandwidth. Without \overcommitname, each receiver allocates $1\times BDP$ worth of credit to the congested sender. With it, each allocates $(BDP + \senderT) / num\_rcvers$. 

\subsection{Simulations}

\label{sec:eval-methodology}

We use network simulation to evaluate \system on large-scale workloads and compare it to other protocol designs. 

We compare \system to 5 baselines: DCTCP~\cite{dctcp}, a widely deployed sender-driven scheme~\cite{microscopic}, Swift~\cite{swift}, a state-of-the-art production sender-driven scheme, Homa~\cite{homa}, because of its near-optimal latency and its use of overcommitment, ExpressPass~\cite{expresspass}, as it employs a hop-by-hop approach to managing credit, and dcPIM~\cite{dcPIM}, as a unique point in the design space as it explicitly matches senders and receivers.    We do not extend Homa with Aeolus~\cite{aeolus} for the reasons discussed in~\cite{homablog-aeolus}. Note that the Homa and dcPIM papers~\cite{homa,dcPIM} already include favorable comparisons to NDP~\cite{ndp}, Aeolus~\cite{aeolus}, PIAS~\cite{pias}, pHost~\cite{phost}, and HPCC~\cite{hpcc}, thus we did not include them in our evaluation.   

We implement \system on ns-2~\cite{altman2012ns}, reuse the original ns-2 DCTCP implementation (using the same parameters as~\cite{pfabric}, scaled to 100Gbps) and the original ns-2 ExpressPass implementation~\cite{github-xpass-sim}, and port the published Homa simulator~\cite{github-homa-sim} to ns-2 using the same parameters as in~\cite{homa} scaled to 100Gbps. The published Homa simulator does not implement the incast optimization~\cite{homa}, which further relies on two-way messages. We use a community Swift simulator, kindly made public \cite{vertigo}, and configure its delay parameters to achieve similar throughput to DCTCP, respecting the guidelines~\cite{swift}. Finally, we use the published dcPIM simulator~\cite{github-dcpim-sim}. \system's simulator is open source and available on GitHub~\cite{sird-simulator}.

\autoref{table:sim_parameters} lists protocol parameter values. DCTCP and Swift use pools of pre-established connections (40 for each host pair).  \system approximates SRPT scheduling like Homa and dcPIM. \system, Homa, and dcPIM use packet spraying, DCTCP and Swift use ECMP. Homa uses 8 network priority levels, dcPIM 3, and \system 2 (\autoref{sec:pacing-scheduling}).   The initial window of DCTCP and Swift is configured at $BDP$ like in~\cite{pfabric}.  The RTT latency is similar to prior work.  

\myparagraph{Topology} We simulate the two-tier leaf-spine topology used in previous work~\cite{homa,dctcp,pfabric,phost,dcPIM} with 144 hosts, connected to 9 top-of-rack switches (16 hosts each) via 4 spine switches, with link speeds of 100Gbps to hosts and 400Gbps to spines.  We simulate switches with infinite buffers, \ie without packet drops (1) to avoid making methodologically complex assumptions as drop rates and thus latency and throughput are very sensitive to switch buffer sizes, organization, and configuration~\cite{microscopic,homablog-aeolus} (see ~\autoref{sec:asictrends}); and (2) to study the intended mode of operation of the protocols that leverage buffering to achieve high link utilization and operate best without drops. The intentional dropping of credit packets to adjust sender rates by ExpressPass is maintained. \system never uses more than a small fraction of the theoretical capacity (\autoref{sec:Throughput-Buffering-Latency}) and is not affected by this setup.

\myparagraph{Workload} Each host operates both as client and server, sending one-way messages according to an open-loop Poisson distribution to uniformly random receivers (all-to-all). We simulate 3 workloads: \textbf{(1) \wthree}: an aggregate of RPC sizes at a Google datacenter~\cite{google-homa-private}; \textbf{(2) \wfour}: a Hadoop workload at Facebook~\cite{inside-fb}; \textbf{(3) \wfive}: a web search application~\cite{pfabric}. We select them to test over a wide range of mean message sizes of 3KB, 125KB, and 2.5MB respectively.


\begin{table}[b!]  \caption{Default simulation parameters for each protocol.} \centering \begin{tabular}{ c | c }
Prot. & Parameters \\
\hline
\multirow{2}{*}{all} & $\bullet$RTT(MSS): 5.5\microsecond intra-rack, 7.5\microsecond inter-rack\\ & $\bullet$$BDP=100KB$; link@100Gbps \\
\hline \multirow{2}{*}{\system} & $\bullet$$\globalB: 1.5 \times BDP$, $\bullet$$\solT: 1 \times BDP$
\\ & $\bullet$$\coreT=1.25 \times BDP$,  $\bullet$$\senderT=0.5 \times BDP$ \\
\hline
\multirow{2}{*}{DCTCP} & $\bullet$Initial window: $1 \times BDP$, $\bullet$g=$0.08$ \\ & $\bullet$Marking ECN Threshold: $1.25 \times BDP$ \\
\hline
\multirow{3}{*}{Swift} & $\bullet$Initial window: $1 \times BDP $ \\ & $\bullet$base\_target: $2 \times RTT$, $\bullet$fs\_range: $5 \times RTT$ \\ & $\bullet$$\hbar: 1.25 \times BDP$, $\bullet$fs\_max: $100$, $\bullet$fs\_min: $0.1$ \\
\hline
\multirow{1}{*}{XPass} & $\bullet\alpha = 1/16$ $\bullet w_{init} = 1/16$ $\bullet loss\_tgt = 1/8$\\
\hline \multirow{1}{*}{Homa} & Same as~\cite{homa} at 100Gbps (incl. priority split) \\  \hline
\multirow{1}{*}{dcPIM} & Same as~\cite{dcPIM} \\
\end{tabular}

\label{table:sim_parameters} \end{table}


We simulate these 3 workloads on 3 traffic configurations, for a total of 9 points of comparison. The configurations are: \textbf{(1) Balanced:} The default configuration described above. We vary the applied load, which does not include protocol-dependent header overheads, from 25\% to 95\% of link capacity. \textbf{(2) Core:} Same as (1) but ToR-Spine links are 200Gbps (2-to-1 oversubscription). Due to uniform message target selection, $128/144 \approx 89\%$ of messages travel via spines, turning the core into the bottleneck. We consequently reduce the load applied by hosts by $\times 0.89*2$ to reflect the network's reduced capacity.   This is not meant to reflect a permanent load distribution, but we hypothesize that it is possible transiently. \textbf{(3) Incast:} We use the methodology of~\cite{hpcc,dcPIM} and combine background traffic with overlay incast traffic: 30 random senders periodically send a 500KB message to a random receiver. Incast traffic represents 7\% of the total load.    

We report goodput (rate of received application payload), total buffering in switches, and message slowdown, defined as the ratio between the measured and the minimum possible latency for each message. In the incast configuration, we exclude incast messages from slowdown results.

\subsubsection{Performance Overview}  \label{sec:Throughput-Buffering-Latency}

\begin{figure*}

\centering \begin{subfigure}{2.3in} \includegraphics[width=\linewidth]{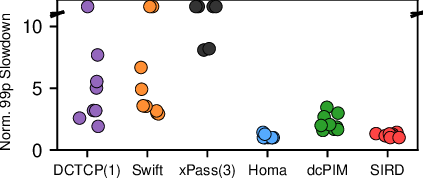} \vspace{-15pt} \caption{Slowdown} \label{fig:scatter-qing-goodput} \end{subfigure} \begin{subfigure}{2.3in} \includegraphics[width=\linewidth]{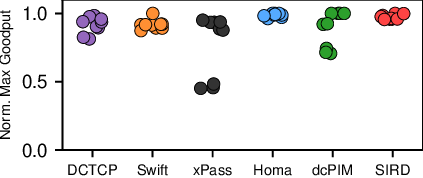} \vspace{-15pt} \caption{Goodput} \label{fig:scatter-slowdown-goodput} \end{subfigure} \begin{subfigure}{2.3in} \includegraphics[width=\linewidth]{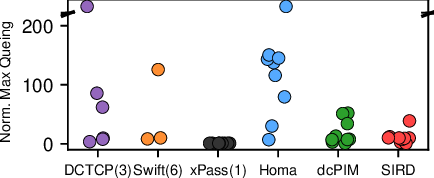} \vspace{-15pt} \caption{Queuing} \label{fig:scatter-qing-slowdown} \end{subfigure}

\caption{ Normalized goodput, queuing, and slowdown across all 9 configurations.  Each metric is normalized based on the best-performing protocol for the given metric and configuration. For queuing and slowdown, lower is better. For goodput, higher is better. Normalized slowdown and buffering are capped at $10\times$ and $200\times$, respectively, and higher values are plotted in the overflow area. The numbers in parentheses show the number of unstable configurations for each protocol which are not plotted. X-axis jitter is added for visibility. Find the data in~\autoref{tbl:normalized}. } \label{fig:eval-scatter} \vspacefigure  \end{figure*}

\autoref{fig:eval-scatter} shows how protocols navigate the trade-offs between throughput, buffering, and latency by plotting their relative performance across all 9 workload-configuration combinations.  The best-performing protocol on each of the 9 scenarios gets a score of $1.0$ (per metric) and the others are normalized to it, so that goodput is always $\le 1.0$ whereas queuing and slowdown are always $\ge 1.0$.  We report the highest achieved goodput and peak queuing over all the levels of applied load.  We report \nineninepct slowdown across all messages of each workload at 50\% applied load which is a level most protocols can deliver in all scenarios. The following figures and calculations do not include cases where a protocol cannot satisfy a specific load level or cannot stop network buffers from growing infinitely.   

Overall, \system is the only protocol that consistently achieves near-ideal scores across all metrics. Specifically, \system causes $\killerbufvshoma \times$ less peak network queuing than Homa and achieves competitive latency and goodput performance. \system outperforms dcPIM in message slowdown, and peak goodput and queuing by $\killerlatvspim\%$, $\killerutilvspim\%$, and $\killerbufvspim\%$ respectively. ExpressPass causes practically zero queuing thanks to its hop-by-hop design, and 88\% less than \system, but \system delivers $\killerlatvsxpass \times$ lower slowdown and $\killerutilvsxpass\%$ more goodput.  Even under full fabric saturation, \system induces at most 0.8MB of ToR queuing in receiver-bottleneck scenarios and 2.3MB in core-bottleneck scenarios. Given a packet buffer capacity of 3.13MB/Tbps (see \autoref{sec:asictrends}), these values correspond to a maximum buffer occupancy of 8\% and 23\%, respectively.

\subsubsection{Congestion Response} \label{sec:Congestion-Response}

\begin{figure}[t]

\centering\includegraphics[scale=0.8]{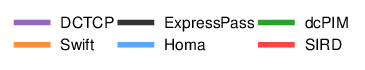} \centering \begin{minipage}{\linewidth} \centering \begin{minipage}{0.325\linewidth} \includegraphics[width=\linewidth]{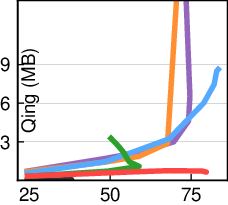} \vspace*{-1.7em} \subcaption{\wthree Balanced} \label{fig:eval-queuing-wka-balanced} \end{minipage} \begin{minipage}{0.325\linewidth} \includegraphics[width=\linewidth]{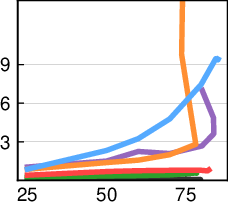} \vspace*{-1.7em} \subcaption{\wfour Balanced} \label{fig:eval-queuing-wkb-balanced} \end{minipage} \begin{minipage}{0.325\linewidth} \includegraphics[width=\linewidth]{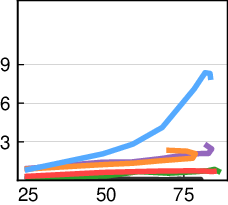} \vspace*{-1.7em} \subcaption{\wfive Balanced} \label{fig:eval-queuing-wkc-balanced} \end{minipage} \end{minipage}

\begin{minipage}{\linewidth} \centering \begin{minipage}{0.325\linewidth} \includegraphics[width=\linewidth]{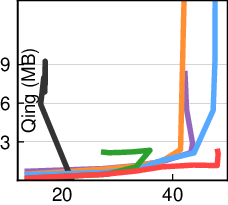} \vspace*{-1.7em} \subcaption{\wthree Core} \label{fig:eval-queuing-wka-core} \end{minipage} \begin{minipage}{0.325\linewidth} \includegraphics[width=\linewidth]{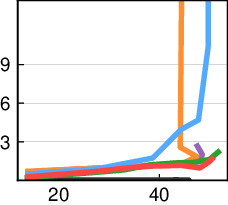} \vspace*{-1.7em} \subcaption{\wfour Core} \label{fig:eval-queuing-wkb-core} \end{minipage} \begin{minipage}{0.325\linewidth} \includegraphics[width=\linewidth]{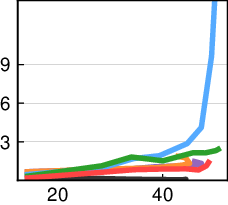} \vspace*{-1.7em} \subcaption{\wfive Core} \label{fig:eval-queuing-wkc-core} \end{minipage} \end{minipage}

\begin{minipage}{\linewidth} \centering \begin{minipage}{0.325\linewidth} \includegraphics[width=\linewidth]{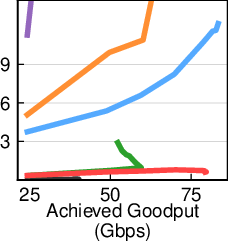} \vspace*{-1.7em} \subcaption{\wthree Incast} \label{fig:eval-queuing-wka-incast} \end{minipage} \begin{minipage}{0.325\linewidth} \includegraphics[width=\linewidth]{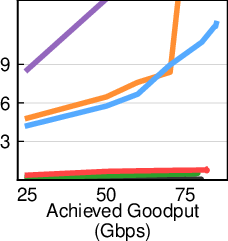} \vspace*{-1.7em} \subcaption{\wfour Incast} \label{fig:eval-queuing-wkb-incast} \end{minipage} \begin{minipage}{0.325\linewidth} \includegraphics[width=\linewidth]{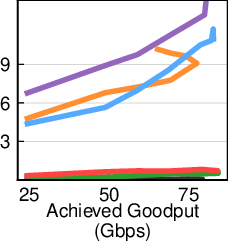} \vspace*{-1.7em} \subcaption{\wfive Incast} \label{fig:eval-queuing-wkc-incast} \end{minipage} \end{minipage}

\caption{ Maximum ToR queuing vs. achieved goodput. Configurations: Balanced (top), Core (middle), Incast (bottom).  }  \label{fig:eval-queuing-merged}   \end{figure}

We now zoom in on how each protocol manages congestion across various levels of applied stress.  \autoref{fig:eval-queuing-merged} plots maximum buffering across ToR switches as a function of achieved goodput for balanced (top), core (middle), and incast (bottom) configurations.  The reported goodput is the mean across all 144 hosts and reflects the rate of message delivery to applications. ToR queuing covers both downlinks and links to aggregation switches (core). 

Overall, \system consistently achieves high goodput while causing minimal buffering across load levels, even under high levels of stress. Through \overcommitname, \system makes effective use of a limited amount of credit and achieves high goodput while reducing the need for packet buffer space by up to $20\times$ compared to Homa's controlled overcommitment.  Further, \system displays a more predictable congestion response compared to dcPIM and ExpressPass.   As expected, based on \autoref{fig:scatter-qing-goodput}, DCTCP and Swift cause meaningful buffering without achieving exceptional goodput.  Mean queuing is qualitatively similar (appendix~\autoref{fig:eval-queuing-merged-mean}). 

dcPIM and ExpressPass sometimes compare favorably to \system but only when the mean message size is not small. They both struggle in terms of achieved goodput and buffering in {\bf \wthree} for all three configurations (left column).  In \wthree, $\approx 99\%$ of messages are smaller than $1\times BDP$ and are responsible for $\approx 40\%$ of the traffic. ExpressPass's behavior is known and discussed in~\cite{expresspass}: more credit than needed may be sent for a small message which may then compete for bandwidth with productive credit of a large message.  For dcPIM, the likely reason as discussed in~\cite{homablog-dcpim} is that to send a small message, a sender has to preempt the transmission of a larger message which can cause the receiver of the latter to remain partially idle.     \system's behavior is consistent across workload message sizes thanks to slight downlink overcommitment which absorbs discontinuities in large message transmission.    

In the core configuration (\autoref{fig:eval-queuing-merged} - middle row), where the core is the bottleneck, we observe that \system's reactive congestion management, despite sharing the same ECN-based mechanism as DCTCP, achieves steadier behavior because it limits the number of bytes in the network. Homa, though lacking an explicit mechanism to regulate queing at the core, does constrain peak queuing below 19MB (6MB/Tbps of switch BW) because it limits outstanding bytes to the aggregate overcommitment level of receivers.  The same applies to dcPIM though queuing is much lower as it does not overcommit.

Last, the incast configuration (bottom row) illustrates the relative advantage of RD schemes over DCTCP and Swift. Homa's results would be different if the incast optimization described in~\cite{homa} was implemented and the workload consisted of request-response pairs instead of one-way messages.   

\begin{figure}[t]  \centering\includegraphics{figures/WKb/tail-all-legend-prot.eps} \begin{minipage}{\linewidth} \begin{subfigure}{0.49\linewidth} \includegraphics[width=\linewidth]{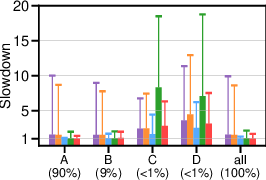} \vspace*{-1.5em} \caption{\wthree Balanced} \end{subfigure} \hfill \begin{subfigure}{0.49\linewidth} \includegraphics[width=\linewidth]{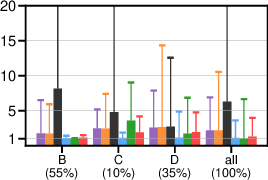} \vspace*{-1.5em} \caption{\wfive Balanced} \label{fig:eval-slowdown-wkc-balanced} \end{subfigure} \end{minipage}

\begin{minipage}{\linewidth} \begin{subfigure}{0.49\linewidth} \includegraphics[width=\linewidth]{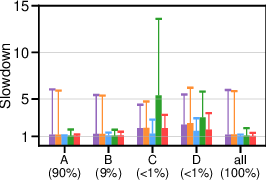} \vspace*{-1.5em} \caption{\wthree Core} \end{subfigure} \hfill \begin{subfigure}{0.49\linewidth} \includegraphics[width=\linewidth]{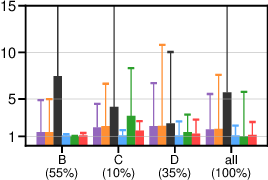} \vspace*{-1.5em} \caption{\wfive Core} \end{subfigure} \end{minipage}

\begin{minipage}{\linewidth} \begin{subfigure}{0.49\linewidth} \includegraphics[width=\linewidth]{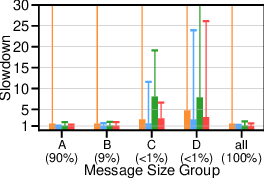} \vspace*{-1.5em} \caption{\wthree Incast} \end{subfigure} \hfill \begin{subfigure}{0.49\linewidth} \includegraphics[width=\linewidth]{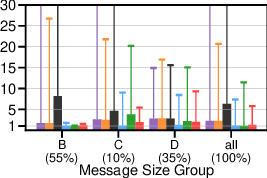} \vspace*{-1.5em} \caption{\wfive Incast} \end{subfigure} \end{minipage}

\caption{Median and \nineninepct slowdown at 50\% offered application load. Each bar group contains the messages in the following size ranges: $0 \leq A < MSS \leq B < 1\times BDP \leq C < 8\times BDP \leq D$. Also shown is the percentage of messages that belong to each group. \wfive has no sub-MSS messages. Protocols that cannot deliver 50\% load are not shown.    } \label{fig:eval-slowdown}  \end{figure}

\subsubsection{Message Latency} \label{sec:message-latency} 

\autoref{fig:eval-slowdown} shows the median and \nineninepct slowdown for different message size ranges across configurations at 50\% applied load (except \wfour results which fall between the other two and can be found in appendix \autoref{fig:eval-slowdown-appendix}).  Across each workload as a whole (rightmost bar cluster "all"), \system is generally on par with Homa, and generally outperforms dcPIM in terms of tail latency. For small, latency-sensitive messages $<BDP$ (100KB - groups A and B), \system, dcPIM, and Homa offer close to hardware latency. Both in aggregate and for small messages, DCTCP and Swift perform an order of magnitude worse at the tail because they cause meaningful buffering without having a bypass mechanism like Homa.

For messages larger than $1\times BDP$ (groups C and D), \system comes closest to near-optimal Homa and strongly outperforms the other protocols. \system achieves up to $4\times$ lower latency than dcPIM in this size range because it does not wait for multi-RTT handshakes before sending a message as discussed in~\autoref{sec:insights}.  Note that \autoref{fig:eval-slowdown-wkc-balanced} is comparable to Figure 3d in~\cite{dcPIM}. Similarly, \system outperforms ExpressPass on latency because it does not take multiple RTTs to capture the full link bandwidth, and because it implements SRPT. 

The differences in latency for protocols other than dcPIM and ExpressPass, which wait before ramping up transmission, can be explained through their \emph{effective} scheduling policies. That is, assuming that a protocol manages to deliver enough messages to achieve 50\% goodput, lowering latency is a matter of appropriately ordering the transmission of messages (assuming no conflicting application concerns). For example, DCTCP achieves equivalent or better latency than Swift in all but the incast scenarios. We argue that this is because Swift is better at fairly sharing bandwidth between messages thanks to its faster-converging control loop.  Latency-wise, fair sharing is inferior to SRPT since it delays individual messages in favor of equitable progress. Along the same lines, \system cannot always match the latency of Homa because it approximates SRPT less faithfully: (1) unlike Homa, a portion of the sender uplink is fair-shared (50\% in this case).     (2) To avoid credit accumulation in congested senders, \overcommitname adjusts per-sender credit bucket sizes equitably. This mostly impacts the latency of group C, as the bucket size may be small due to sender congestion and not grow during the lifetime of a message smaller than $8\times BDP$. Consequently, compared to near-optimal Homa in the Balanced configuration, \system's \nineninepct slowdown in group C is $1.85\times$ and $2.68\times$ higher at 50\% and 70\% application load respectively, while it outperforms the other protocols. 

\begin{figure}[t]   \begin{minipage}{\linewidth} \begin{subfigure}{0.49\linewidth} \includegraphics[width=\linewidth]{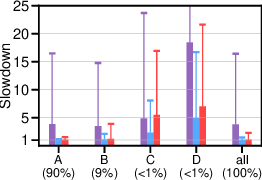} \vspace*{-1.5em} \caption{\wthree Balanced} \end{subfigure} \hfill \begin{subfigure}{0.49\linewidth} \includegraphics[width=\linewidth]{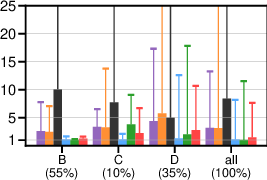} \vspace*{-1.5em} \caption{\wfive Balanced} \label{fig:eval-slowdown-wkc-balanced-hl} \end{subfigure} \end{minipage}

\caption{Median and \nineninepct slowdown at 70\% offered application load.     Protocols that cannot deliver 70\% load are not shown. Legend same as in~\autoref{fig:eval-slowdown}.  } \label{fig:eval-slowdown-high}  \end{figure}

\autoref{fig:eval-slowdown-high} shows slowdown under a higher applied application load of 70\% for the protocols that can deliver it. At this load level, message scheduling becomes more important and Homa's near-optimal SRPT implementation sees its relative performance improve in some cases.

In summary, \system delivers messages with low latency, outperforms dcPIM, ExpressPass, DCTCP, and Swift, and is competitive with Homa, which nearly optimally approximates SRPT~\cite{homa}.

\subsubsection{Sensitivity Analysis} \label{sec:eval-sensitivity}

In this section, we explore \system's sensitivity to its key parameters: $\globalB$, $\senderT$, and $\solT$, as well as to the availability of switch priority queues. \system's sensitivity to $\coreT$ is the same as DCTCP's to $K$~\cite{dctcp}.

\begin{figure} \begin{subfigure}{0.695\linewidth} \includegraphics[width=\columnwidth]{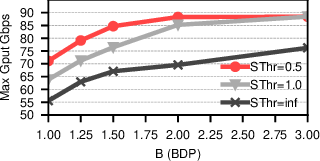} \end{subfigure} \hfill \begin{subfigure}{0.295\linewidth} \includegraphics[width=\columnwidth]{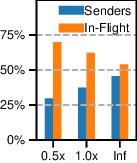} \end{subfigure}

\caption{Left: Max goodput across values of $\globalB$ and $\senderT$. Right: Credit location at max goodput as a function of $\senderT$. $B=1.5\times BDP$. "In-flight" is \grant and \data. Credit at receivers is low at this load.  } \label{fig:eval-sensitivity} \vspacefigure  \end{figure}

\myparagraph{\Overcommitname} \autoref{equation:htr} introduced the linear steady-state relationship between $\globalB$ and $\senderT$, the two key parameters of \overcommitname.  \autoref{fig:eval-sensitivity} (left) shows the measured maximum achieved goodput in Balanced \wfive as a function of $\globalB$ and $\senderT$. 

We observe that the presence of \overcommitname increases achievable goodput by \twiddle $25\%$, confirming that the introduced sender-informed mechanism is necessary to fully utilize the network with a limited amount of credit, and thus low queuing. When the mechanism is disabled ($\senderT=\inf$), there is stranded and unused credit at congested senders, which prevents receivers from achieving the full rate.   \autoref{fig:eval-sensitivity} (right) shows where in the topology credit is under max goodput. Lower $\senderT$ values reduce the credit accumulation at senders, and improves throughput by increasing the number of in-flight \grant and \data packets. 

With the mechanism enabled, the curves asymptotically converge to the same maximum goodput of 90Gbps. Reaching the plateau with a lower value of $B$ also demands lowering $\senderT$ to reduce credit stranded at senders (\autoref{equation:htr}).   Queuing increases with $\globalB$ as in~\autoref{fig:frontier} and remains stable when varying $\senderT$.  We selected $\globalB=1.5\times BDP$ rather than $\globalB=2.0\times BDP$ for our experiments because it halves the maximum buffering caused by scheduled packets from $1\times$ to $0.5\times BDP$ ($=\globalB-BDP$).     We do not configure $\senderT$ lower than $0.5\times BDP$ as we deem it unrealistic to implement in software, as it may cause unwanted marking due to batch credit arrivals. The minimum correct value for this topology is $0.15\times BDP$, as per the guidelines~\cite{dctcp}. 

\begin{figure} 

\begin{subfigure}{0.49\linewidth} \includegraphics[width=\linewidth]{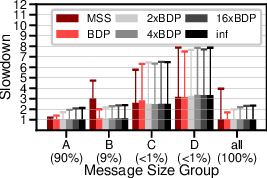} \end{subfigure} \hfill \begin{subfigure}{0.49\linewidth} \includegraphics[width=\linewidth]{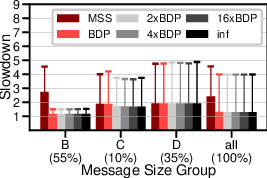} \end{subfigure}

\caption{ Slowdown as a function of \solT for \wthree (left) and \wfive (right) at 50\% application load in the balanced configuration;  } \label{fig:unsched-unsol} \vspacefigure \end{figure}

\myparagraph{Unscheduled transmissions} \system allows the unscheduled transmission of the $BDP$ prefix of messages $\le\solT$.  \autoref{fig:unsched-unsol} explores the sensitivity to this parameter using workloads \wthree and \wfive. Setting $\solT = MSS$ meaningfully increases median and tail latency for messages with sizes in $[MSS,BDP]$ while higher values offer no appreciable net benefit. This also holds at 70\% load. Increasing $\solT >> BDP$ in \wfive has little impact on max and mean ToR buffering, which range within $[696,747]$KB and $[363,378]$KB respectively.  In contrast, because $99\%$ of messages in \wthree are unscheduled, max and mean ToR buffering scale from $501$KB to $1016$KB and $220$KB to $435$KB, respectively, as \solT increases.  It is worth noting that saturating the fabric's bandwidth, which is when queuing is maximized, is far less likely with high-request-rate workloads like \wthree.

The default value of $\solT=1\times BDP$ is a satisfying compromise as large values do not yield latency benefits yet can unnecessarily expose the fabric to coordinated traffic bursts (\eg incast). We confirm this by running \wfive under the incast configuration which has a high concentration of $5\times BDP$ message bursts. We observe the following performance degradation when comparing $\solT=4\times BDP$ to $\solT=16\times BDP$: Overall \nineninepct slowdown increases by 34\% while maximum and mean ToR queuing increases by $5.7\times$ and $80\%$ respectively. These results justify our decision to introduce a size threshold above which messages are entirely scheduled.

In deployments with severe incast consisting of messages smaller than $BDP$, \solT can be configured to be smaller than $MSS$. If, for example, 100 hosts concurrently send a $0.5\times BDP$ message to a single receiver, the latter will effectively receive $100$ small packets requesting credit. The receiver will immediately start issuing credit based on its policy and ramp-up to full bandwidth utilization one RTT after the senders transmitted. In TCP terms, this specific scenario is roughly equivalent to each sender starting with a window of one MSS, and ramping to the correct window value just one RTT later.

\begin{figure} 

\begin{subfigure}{0.49\linewidth} \includegraphics[width=\linewidth]{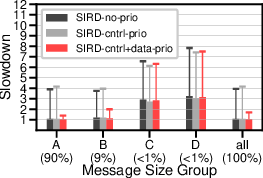} \end{subfigure} \hfill \begin{subfigure}{0.49\linewidth} \includegraphics[width=\linewidth]{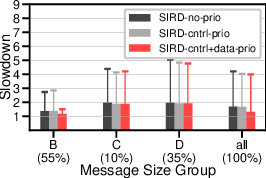} \end{subfigure}

\caption{ Slowdown as a function of priority use for \wthree (left) and \wfive (right) at 50\% application load in the balanced configuration;  } \label{fig:unsched-prio} \vspacefigure \end{figure}

\myparagraph{Use of switch priority queues} \system may use a second 802.1p priority level for control packets and/or unscheduled \data. \autoref{fig:unsched-prio} shows how the use of priorities impacts message slowdown for \wthree and \wfive in terms of (i) possible degradation when switch priority queues are unavailable, and (ii) the benefit of prioritizing unscheduled \data.

We observe that, across message sizes, median slowdown is largely unaffected and tail slowdown benefits from high priority transmission in some cases. Small messages benefit because they bypass the small queues \system forms ($\approx 1\times BDP$ per link at the \nineninepct and $\approx 0.1\times BDP$ on average). This is more evident in \wthree where 99\% of messages are unscheduled.  Sensitivity to priorities is similar at 70\% load (not shown), where, compared to 50\% load, group A's tail slowdown benefits 5\% more from \data prioritization and group B's 6\% less.    

Maximum goodput increases by $1\%$ and $2.4\%$ with control packet prioritization for \wthree and \wfive respectively, as credit is delivered slightly more predictably. Queuing in the network is also insensitive to the use of priorities. Given \system's low sensitivity to the availability of switch priority queues, we argue that it can be deployed without this dependency with little sacrifice.  This differentiates \system from Homa, which is designed around the use of priorities to bypass long in-network queues, and dcPIM, which strongly benefits from tight control packet delivery due to its semi-synchronous nature. 

\section{Related Work} \label{sec:related}

The topic of datacenter congestion control has been extensively explored due to the increasing IO speeds, the \microsecond-scale latencies, emerging programmable hardware, RPC workloads, and host-centric concerns~\cite{dctcp,hull,pfabric,on-ramp,swift,hpcc,timely,fastpass,pl2,homa,phost,pias,ndp,ppt,harmony,expresspass,snap,erpc,r2p2,dcqcn,powertcp,pase,d2tcp,hostcc,breakwater,ccreview,pdq,d3,tlt,fastlane,cp, poseidon,sfc,bolt,bfp}.  

SD schemes like DCQCN~\cite{dcqcn}, Timely~\cite{timely}, HPCC~\cite{hpcc}, Bolt~\cite{bolt}, and PTT~\cite{ppt} use network signals in the form of ECN~\cite{dctcp, dcqcn, d2tcp}, delay~\cite{timely,swift}, in-band telemetry~\cite{hpcc,poseidon,fasttune}, and packet drops~\cite{pfabric} to adjust sending windows~\cite{dctcp, swift, pase,hpcc} or rates~\cite{timely,pdq,d3}. Some operate end-to-end with no or trivial switch support~\cite{timely, swift, dctcp, ppt} while others leverage novel but non-universal switch features to improve feedback quality~\cite{hpcc,poseidon,fasttune} or to tighten reaction times~\cite{bolt,fastlane,cp}.    

RD protocols attempt to avoid congestion altogether by proactively scheduling packet transmissions~\cite{aeolus,ndp,pl2,homa,dcPIM,expresspass,fastpass,phost,flexpass,harmony} and have already been discussed (\autoref{sec:back}). Orthogonally to the scope of this paper, FlexPass~\cite{flexpass} enables fair bandwidth sharing among proactive and reactive protocols. EQDS~\cite{eqds} enables the coexistence of traditional protocols like DCTCP and RDMA by moving all queuing interactions to end-hosts and using an RD approach to admit packets into the network fabric.        

Previous work has followed different approaches regarding the desired flow/message scheduling policy. The traditional approach is to aim for fair bandwidth allocation at the flow level~\cite{dctcp,swift,timely,dcqcn}.  Alternatively, D\textsuperscript{2}TCP, D\textsuperscript{3}, and Karuna~\cite{karuna} prioritize flows based on deadlines communicated by higher-level services. A recent line of work~\cite{homa,phost,pfabric,pias,aeolus}, which includes both proactive and reactive protocols, attempts to approximate SRPT or SJF (shorted job first) scheduling in the fabric which minimizes average latency~\cite{pfabric,srptproof}. \system can approximate a variety of policies at receivers and senders and avoids the need for fabric policies as it keeps in-network queuing minimal. 


\section{Conclusion} \label{sec:conclusion}

\system takes a holistic approach to datacenter congestion management by tackling the fundamental tension of receiver-driven designs, which is the management of shared links. \system applies proactive scheduling to exclusive receiver downlinks, which are statistically the most congested, and leverages sender and fabric congestion feedback to reactively allocate the bandwidth of shared links. Compared to existing designs, \system manages to simultaneously deliver high link utilization with hardly any network buffering, and low message latency. \system does so without assuming advanced switch ASIC features nor switch priority queues, which makes it compatible with existing and heterogeneous datacenter networks. We implemented a prototype of \system in software, on the Caladan stack, and showed that it can efficiently allocate credit while delivering 100Gbps.

\section*{Acknowledgements}

The authors thank John Ousterhout for numerous discussions on the topic of receiver-driven protocols. We thank our shepherd Srikanth Sundaresan, Nate Foster, Katerina Argyraki, Haitham Al Hassanieh, James Larus, Charly Castes, Neelu Kalani, Boris Pismenny, Rui Yang, Mahyar Emami, Sahand Kashani, Mia Primorac, Francois Costa, and all anonymous reviewers for their insightful comments.

This work was funded in part by the Microsoft-EPFL Joint Research Center and used CloudLab~\cite{cloudlab} for 100GbE experiments. Ryan Kosta participated in EPFL's Excellence Research Internship Program.

\newpage \balance \bibliographystyle{plain} \bibliography{gen-abbrev,dblp,misc}


\newpage \appendix

\section{Appendix} \label{app:appendix}


\begin{table}[h]

\caption{ ASIC bisection bandwidth (in Tbps) and buffer sizes (in MB). Note that buffer architectures differ regarding the extent to which they are shared between ports. } \label{tbl:switches} \fontsize{8}{10}\selectfont \centering \begin{tabular}{ | c| c | c | c |  } \hline
&  ASIC/Model  & BW  & Buffer  \\
\hline
\multirow{14}{*}{\rotatebox[origin=c]{90}{Broadcom}} &    Trident+ & 0.64 & 9  \\

& Trident2  & 1.28 & 12  \\

& Trident2+  & 1.28 & 16  \\

& Trident3-X4  & 1.7 & 32  \\

& Trident3-X5  & 2 & 32  \\

& Tomahawk  & 3.2 & 16  \\

& Trident3-X7  & 3.2 & 32  \\

&  Tomahawk 2  & 6.4 & 42  \\

&  Tomahawk 3 BCM56983  & 6.4 & 32  \\

& Tomahawk 3 BCM56984  & 6.4 & 64  \\

& Tomahawk 3 BCM56982  & 8 & 64  \\

&  Tomahawk 3  & 12.8 & 64  \\

&  Trident4 BCM56880  & 12.8 & 132  \\

&  Tomahawk 4  & 25.6 & 113  \\
\hline
\multirow{8}{*}{\rotatebox[origin=c]{90}{nVidia}}& Spectrum SN2100  & 1.6 & 16  \\

& Spectrum SN2410  & 2 & 16  \\

&        Spectrum SN2700  & 3.2 & 16  \\

&        Spectrum SN3420  & 2.4 & 42  \\

&        Spectrum SN3700  & 6.4 & 42  \\

&        Spectrum SN3700C  & 3.2 & 42  \\

&        Spectrum SN4600C  & 6.4 & 64  \\

&        Spectrum SN4410  & 8 & 64  \\

&        Spectrum SN4600  & 12.8 & 64  \\

&        Spectrum SN4700  & 12.8 & 64  \\

&        Spectrum SN5400  & 25.6 & 160  \\

&        Spectrum SN5600  & 51.2 & 160  \\
\hline   

\end{tabular} 

\end{table}

\vspace*{20em}

\begin{figure} \centering\includegraphics{figures/WKb/tail-all-legend-prot.eps}

\begin{subfigure}{0.7\linewidth} \includegraphics[width=\linewidth]{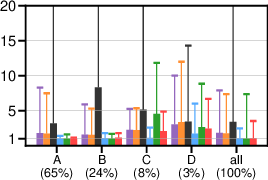} \vspace*{-1.5em} \caption{\wfour Balanced} \end{subfigure}

\begin{subfigure}{0.7\linewidth} \includegraphics[width=\linewidth]{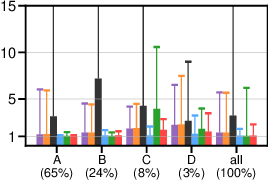} \vspace*{-1.5em} \caption{\wfour Core} \end{subfigure}

\begin{subfigure}{0.7\linewidth} \includegraphics[width=\linewidth]{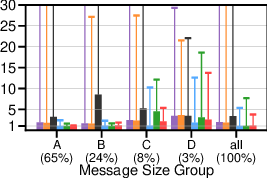} \vspace*{-1.5em} \caption{\wfour Incast} \end{subfigure}

\caption{Median and \nineninepct slowdown at 50\% offered application load. Each bar group contains the messages in the following size ranges: $0 \leq A < MSS \leq B < 1\times BDP \leq C < 8\times BDP \leq D$. } \label{fig:eval-slowdown-appendix} \end{figure}

\begin{table*}[th] \vspace{2em} \centering \begin{tabular}{|l|rrr|rrr|rrr|rr|} \toprule
Config & \multicolumn{3}{|c|}{Default} & \multicolumn{3}{|c|}{Core} & \multicolumn{3}{|c|}{Incast} & mean & range\\

Wload & WKa & WKb & WKc & WKa & WKb & WKc & WKa & WKb & WKc & &  \\
\midrule

DCTCP  & 7.69 & 3.19 & 1.91 & 5.02 & 3.18 & 2.58 & unstable & 18.88 & 5.54 & 6.0 & 16.97 \\
Swift  & 6.68 & 2.98 & 2.92 & 4.92 & 3.16 & 3.53 & 53.5 & 10.23 & 3.57 & 10.17 & 50.58 \\
ExpressPass  & unstable & 14.94 & 8.09 & unstable & 15.9 & 11.37 & unstable & 12.61 & 8.2 & 11.85 & 7.81 \\
Homa  & 1.0 & 1.0 & 1.0 & 1.0 & 1.0 & 1.0 & 1.0 & 1.42 & 1.27 & 1.08 & 0.42 \\
dcPIM  & 1.67 & 2.98 & 1.84 & 1.58 & 3.45 & 2.68 & 1.66 & 2.05 & 1.98 & 2.21 & 1.87 \\
SIRD  & 1.32 & 1.43 & 1.1 & 1.15 & 1.27 & 1.18 & 1.3 & 1.0 & 1.0 & 1.19 & 0.43 \\

\bottomrule
\multicolumn{12}{|c|}{Normalized 99th percentile slowdown of all messages at 50\% load.} \\
\toprule DCTCP  & 0.9 & 0.98 & 0.98 & 0.9 & 0.94 & 0.94 & 0.81 & 0.83 & 0.96 & 0.92 & 0.17 \\
Swift  & 0.89 & 0.92 & 0.91 & 0.87 & 0.92 & 0.89 & 0.92 & 1.0 & 0.92 & 0.92 & 0.13 \\
ExpressPass  & 0.46 & 0.93 & 0.93 & 0.45 & 0.89 & 0.88 & 0.48 & 0.94 & 0.95 & 0.77 & 0.5 \\
Homa  & 1.0 & 1.0 & 0.97 & 0.99 & 0.96 & 0.99 & 1.0 & 0.99 & 0.98 & 0.99 & 0.04 \\
dcPIM  & 0.71 & 0.92 & 1.0 & 0.74 & 1.0 & 1.0 & 0.71 & 0.92 & 1.0 & 0.89 & 0.29 \\
SIRD  & 0.96 & 0.97 & 0.98 & 1.0 & 0.98 & 0.96 & 0.96 & 0.96 & 1.0 & 0.97 & 0.04 \\
  \bottomrule
\multicolumn{12}{|c|}{Normalized maximum goodput across applied load levels.} \\
\toprule DCTCP  & unstable & 85.81 & 9.71 & 3.68 & 62.13 & 7.94 & unstable & unstable & 260.51 & 71.63 & 256.83 \\
Swift  & unstable & unstable & 8.41 & unstable & unstable & 10.12 & unstable & unstable & 125.77 & 48.1 & 117.36 \\
ExpressPass  & 1.0 & 1.0 & 1.0 & unstable & 1.0 & 1.0 & 1.0 & 1.0 & 1.0 & 1.0 & 0.0 \\
Homa  & 137.45 & 115.99 & 30.18 & 7.06 & 434.71 & 79.62 & 143.38 & 150.57 & 145.29 & 138.25 & 427.65 \\
dcPIM  & 52.07 & 7.76 & 3.1 & 1.04 & 51.24 & 13.09 & 34.25 & 7.18 & 7.41 & 19.68 & 51.03 \\
SIRD  & 12.05 & 9.91 & 2.68 & 1.0 & 38.9 & 7.53 & 9.37 & 10.03 & 10.27 & 11.3 & 37.9 \\
  \bottomrule
\multicolumn{12}{|c|}{Normalized maximum ToR queuing across applied load levels.} \\
\bottomrule \end{tabular}

\caption{Normalized data used in \autoref{fig:eval-scatter}. Performance is normalized to the best performing protocol on each experiment. Experiments in which the protocol is unable to deliver the specified throughput or network queuing grows infinitely (unstable) are excluded from the calculation of mean and range.} \label{tbl:normalized}

\end{table*}

\begin{table*}[th] \vspace{2em} \centering \begin{tabular}{|l|rrr|rrr|rrr|rr|} \toprule
Config & \multicolumn{3}{|c|}{Default} & \multicolumn{3}{|c|}{Core} & \multicolumn{3}{|c|}{Incast} & mean & range\\

Wload & WKa & WKb & WKc & WKa & WKb & WKc & WKa & WKb & WKc & &  \\
\midrule

DCTCP  & 9.92 & 7.9 & 6.91 & 5.97 & 5.71 & 5.54 & unstable & 71.06 & 32.06 & 18.13 & 65.52 \\
Swift  & 8.61 & 7.37 & 10.56 & 5.85 & 5.68 & 7.58 & 68.21 & 38.51 & 20.69 & 19.23 & 62.53 \\
ExpressPass  & unstable & 36.94 & 29.26 & unstable & 28.56 & 24.42 & unstable & 47.45 & 47.45 & 35.68 & 23.03 \\
Homa  & 1.29 & 2.47 & 3.62 & 1.19 & 1.8 & 2.15 & 1.27 & 5.36 & 7.37 & 2.95 & 6.18 \\
dcPIM  & 2.16 & 7.37 & 6.66 & 1.88 & 6.2 & 5.76 & 2.11 & 7.7 & 11.48 & 5.7 & 9.6 \\
SIRD  & 1.7 & 3.53 & 3.99 & 1.37 & 2.28 & 2.54 & 1.65 & 3.76 & 5.79 & 2.96 & 4.42 \\

\bottomrule
\multicolumn{12}{|c|}{99th percentile slowdown of all messages at 50\% load.} \\
\toprule DCTCP  & 74.65 & 83.85 & 83.95 & 43.63 & 48.54 & 47.42 & 67.8 & 70.53 & 80.74 & 66.79 & 40.32 \\
Swift  & 74.52 & 78.16 & 78.69 & 42.16 & 47.41 & 45.15 & 76.8 & 85.46 & 77.57 & 67.32 & 43.3 \\
ExpressPass  & 38.04 & 79.68 & 80.42 & 21.78 & 45.83 & 44.42 & 40.26 & 80.1 & 80.1 & 56.74 & 58.64 \\
Homa  & 83.39 & 85.23 & 83.55 & 47.73 & 49.73 & 50.02 & 83.39 & 84.87 & 82.79 & 72.3 & 37.5 \\
dcPIM  & 58.94 & 78.42 & 86.03 & 35.91 & 51.68 & 50.59 & 59.61 & 79.02 & 84.26 & 64.94 & 50.12 \\
SIRD  & 79.74 & 82.27 & 84.71 & 48.27 & 50.47 & 48.75 & 79.7 & 81.98 & 84.13 & 71.11 & 36.44 \\
  \bottomrule
\multicolumn{12}{|c|}{Maximum goodput across applied load levels (Gbps).} \\
\toprule DCTCP  & unstable & 7.0 & 2.7 & 8.3 & 2.67 & 1.46 & unstable & unstable & 21.06 & 7.2 & 19.6 \\
Swift  & unstable & unstable & 2.33 & unstable & unstable & 1.87 & unstable & unstable & 10.17 & 4.79 & 8.3 \\
ExpressPass  & 0.06 & 0.08 & 0.28 & unstable & 0.04 & 0.18 & 0.08 & 0.08 & 0.08 & 0.11 & 0.24 \\
Homa  & 8.63 & 9.46 & 8.37 & 15.91 & 18.68 & 14.68 & 12.17 & 12.17 & 11.75 & 12.42 & 10.31 \\
dcPIM  & 3.27 & 0.63 & 0.86 & 2.34 & 2.2 & 2.41 & 2.91 & 0.58 & 0.6 & 1.76 & 2.69 \\
SIRD  & 0.76 & 0.81 & 0.75 & 2.26 & 1.67 & 1.39 & 0.79 & 0.81 & 0.83 & 1.12 & 1.51 \\
  \bottomrule
\multicolumn{12}{|c|}{Maximum ToR queuing across applied load levels (MB).} \\
\bottomrule \end{tabular}

\caption{Raw data used in \autoref{fig:eval-scatter}. Experiments in which the protocol is unable to deliver the specified throughput or network queuing grows infinitely (unstable) are excluded from the calculation of mean and range.} \label{tbl:notnormalized}

\end{table*}


\begin{figure*}[]

\centering\includegraphics{figures/WKb/tail-all-legend-prot.eps}  \centering \begin{minipage}{\linewidth} \centering \begin{minipage}{0.2\linewidth} \includegraphics[width=\linewidth]{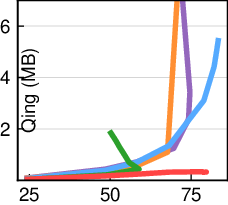} \subcaption{\wthree Balanced} \label{fig:eval-queuing-wka-balanced-mean} \end{minipage} \begin{minipage}{0.2\linewidth} \includegraphics[width=\linewidth]{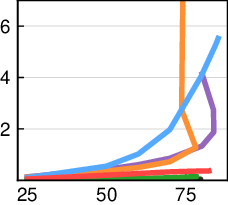} \subcaption{\wfour Balanced} \label{fig:eval-queuing-wkb-balanced-mean} \end{minipage} \begin{minipage}{0.2\linewidth} \includegraphics[width=\linewidth]{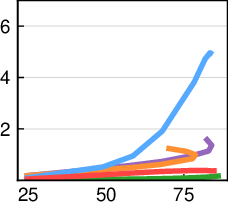} \subcaption{\wfive Balanced} \label{fig:eval-queuing-wkc-balanced-mean} \end{minipage} \end{minipage}

\begin{minipage}{\linewidth} \centering \begin{minipage}{0.2\linewidth} \includegraphics[width=\linewidth]{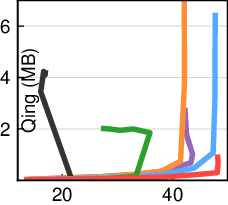} \subcaption{\wthree Core} \label{fig:eval-queuing-wka-core-mean} \end{minipage} \begin{minipage}{0.2\linewidth} \includegraphics[width=\linewidth]{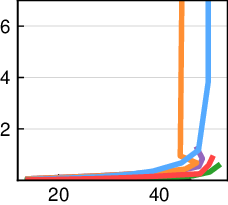} \subcaption{\wfour Core} \label{fig:eval-queuing-wkb-core-mean} \end{minipage} \begin{minipage}{0.2\linewidth} \includegraphics[width=\linewidth]{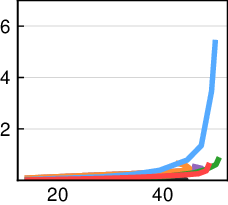} \subcaption{\wfive Core} \label{fig:eval-queuing-wkc-core-mean} \end{minipage} \end{minipage}

\begin{minipage}{\linewidth} \centering \begin{minipage}{0.2\linewidth} \includegraphics[width=\linewidth]{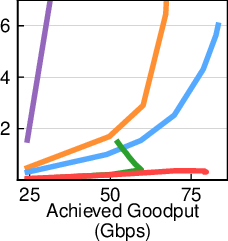} \subcaption{\wthree Incast} \label{fig:eval-queuing-wka-incast-mean} \end{minipage} \begin{minipage}{0.2\linewidth} \includegraphics[width=\linewidth]{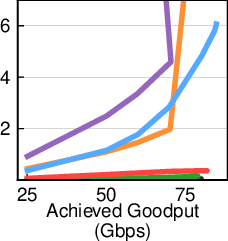} \subcaption{\wfour Incast} \label{fig:eval-queuing-wkb-incast-mean} \end{minipage} \begin{minipage}{0.2\linewidth} \includegraphics[width=\linewidth]{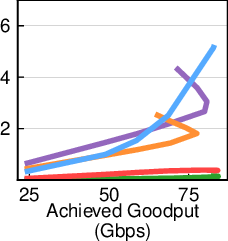} \subcaption{\wfive Incast} \label{fig:eval-queuing-wkc-incast-mean} \end{minipage} \end{minipage}

\caption{Mean ToR queuing vs. achieved goodput. Configurations: Balanced (top), Core (middle), Incast (bottom).}   \label{fig:eval-queuing-merged-mean} \end{figure*}



\end{document}